# A Comparative Study on Spin-Orbit Torque Efficiencies from W/ferromagnetic and W/ferrimagnetic Heterostructures

Ting-Chien Wang, Tian-Yue Chen[†], Chun-Te Wu, Hung-Wei Yen, and Chi-Feng Pai[*]

*Department of Materials Science and Engineering, National Taiwan University, Taipei 10617, Taiwan*

It has been shown that W in its resistive form possesses the largest spin-Hall ratio among all heavy transition metals, which makes it a good candidate for generating efficient dampinglike spin-orbit torque (DL-SOT) acting upon adjacent ferromagnetic or ferrimagnetic (FM) layer. Here we provide a systematic study on the spin transport properties of W/FM magnetic heterostructures with the FM layer being ferromagnetic $Co_{20}Fe_{60}B_{20}$ or ferrimagnetic $Co_{63}Tb_{37}$ with perpendicular magnetic anisotropy. The DL-SOT efficiency $|\xi_{DL}|$, which is characterized by a current-induced hysteresis loop shift method, is found to be correlated to the microstructure of W buffer layer in both $W/Co_{20}Fe_{60}B_{20}$ and $W/Co_{63}Tb_{37}$ systems. Maximum values of $|\xi_{DL}| \approx 0.144$ and $|\xi_{DL}| \approx 0.116$ are achieved when the W layer is partially amorphous in the $W/Co_{20}Fe_{60}B_{20}$ and $W/Co_{63}Tb_{37}$ heterostructures, respectively. Our results suggest that the spin Hall effect from resistive phase of W can be utilized to effectively control both ferromagnetic and ferrimagnetic layers through a DL-SOT mechanism.

[†] Email: r05527067@ntu.edu.tw
[*] Email: cfpai@ntu.edu.tw

# I. INTRODUCTION

The spin Hall effect (SHE) [1-3] is the phenomenon of conduction electrons with different spin orientations being deflected to different transverse directions due to spin-orbit interactions in materials. Among all classical materials (excluding emergent materials systems such as topological insulators), 5d transition metals (TM) like Pt [4], Ta [5], and W [6] are reported to have significant SHE. In order to utilize the SHE from these materials, transition metal/ferromagnetic metal (TM/FM) bilayer heterostructures are typically employed to observe SHE-induced magnetization switching [5,7], domain-wall motion [8,9], and dynamics [10,11]. In a TM/FM bilayer structure, the SHE-induced transverse spin current $J_s$ that quantifies the spins being absorbed by the FM layer can be expressed as $J_s = (\hbar/2e) T_{int}^{TM/FM} \theta_{SH}^{TM} J_e^{TM}$, where $\theta_{SH}^{TM}$ is the internal spin Hall ratio of the TM layer and $J_e^{TM}$ represents the longitudinal charge current density flowing in the TM layer. $T_{int}^{TM/FM}$ represents the spin transparency at the TM/FM interface [12,13] (note that $T_{int}^{TM/FM} = 1$ for a perfect transmission), which is related to the spin-mixing conductance [14,15]. The transmitted spin angular momentum from $J_s$ can be transferred to the magnetic moments in FM through a spin-transfer torque mechanism [16]. The final effect is therefore a dampinglike spin-orbit torque (DL-SOT) acting upon the FM layer, and the DL-SOT efficiency of a TM/FM bilayer heterostructure can be expressed as $\xi_{DL} \equiv (2e/\hbar) J_s / J_e^{TM} = T_{int}^{TM/FM} \theta_{SH}^{TM}$ [12].

For TMs, W has been reported to possess the largest internal spin Hall ratio and DL-SOT efficiency at room temperature, $|\xi_{DL}| \approx 0.30$ [17]. The efficacy of the SHE from W has been demonstrated and characterized through DL-SOT switching measurements [6,18,19], harmonic voltage methods [20], spin-pumping measurements [21], spin-Hall magnetoresistance measurements [22-24], and optical approaches [25,26]. The strength of the SHE in W is also known to be strongly phase dependent: The resistive β-phase or amorphous phase has been experimentally shown to have greater SHE and DL-SOT efficiency while compare to the conductive α-phase or crystalline phase [6,22,26]. In this work, we provide a detailed systematic study on the DL-SOT efficiencies $|\xi_{DL}|$ from two series of W/FM magnetic heterostructures using a newly-developed hysteresis loop shift measurement method [27]. The FM layers that we employed are perpendicularly-magnetized ferromagnetic $Co_{20}Fe_{60}B_{20}$ and ferrimagnetic $Co_{63}Tb_{37}$. Ferromagnetic Co-Fe-B is a classical FM layer for magnetic tunnel junctions [28,29] and SHE three-terminal device applications [5,30], whereas ferrimagnetic Co-Tb and Co-Gd alloys have been adopted to study SOT behavior near the compensation point of the FM layer [31-33]. By characterizing $|\xi_{DL}|$ from both W/$Co_{20}Fe_{60}B_{20}$ and W/$Co_{63}Tb_{37}$ structures, the comparative study allows us to examine different possible spin transport scenarios (different $T_{int}^{W/FM}$) across two different W/FM interfaces. We find that $|\xi_{DL}|$ depends on the thickness and the microstructure of buffer layer W in both systems. $|\xi_{DL}|$ can reach ~0.144 and ~0.116 for W/$Co_{20}Fe_{60}B_{20}$ and W/$Co_{63}Tb_{37}$ heterostructures, respectively, when the W buffer layer is thin (≤4 nm), resistive, and partially amorphous. Our results suggest that resistive W can serve as an efficient DL-SOT source and the corresponding $T_{int}^{W/FM}$ is comparable for FM being either ferromagnetic or ferrimagnetic.

## II. HETEROSTRUCTURES PREPARATION

Magnetic heterostructures of this work are deposited in a high vacuum sputtering chamber with base pressure of $3\times10^{-8}$ Torr. We use dc (rf) magnetron sputtering with 3 mTorr (10 mTorr) of Ar working pressure for depositions of metallic (oxide) layers. Multilayer stack heterostructures W($t_\mathrm{W}$)/Co$_{20}$Fe$_{60}$B$_{20}$(1.4)/Hf(0.5)/MgO(2)/Ta(2) (numbers in the parenthesis are in nanometers) and W($t_\mathrm{W}$)/Co$_{63}$Tb$_{37}$(10)/Ta(2) are deposited onto thermally-oxidized silicon substrates, with $t_\mathrm{W}$ ranges from 1 nm to 16 nm. Co$_{20}$Fe$_{60}$B$_{20}$ (Co-Fe-B) is directly sputtered from a single target while Co$_{63}$Tb$_{37}$ (Co-Tb) is prepared by co-sputtering of a Co and a Tb target. Samples from the W/Co-Fe-B series are annealed at 300 ℃ for 1 hour in high vacuum to induce perpendicular magnetic anisotropy (PMA) of the Co-Fe-B layer. The Hf(0.5) insertion layer serves as a PMA enhancing layer and should be mostly oxidized therefore having minimal effects on transport properties [34]. No further heat treatment is needed to obtain PMA for the W/Co-Tb series since ferrimagnet Co-Tb possesses bulk PMA [35,36]. The Ta(2) capping layers protect the heterostructures beneath from oxidation. To perform transport measurements, heterostructures are made into micron-sized Hall-bar devices with lateral dimensions of $5\,\mu\mathrm{m}\times60\,\mu\mathrm{m}$ and $10\,\mu\mathrm{m}\times60\,\mu\mathrm{m}$ by photolithography.

## III. RESULTS FROM W/Co-Fe-B HETEROSTRUCTURES

### A. Magnetic properties and microstructures

We first characterize magnetic anisotropy of W/Co-Fe-B heterostructures with magneto-optical Kerr effect (MOKE). As shown in Fig. 1(a) and (b), the out-of-plane hysteresis loops of W/Co-Fe-B with thin (4 nm) W buffer layer and thick (14 nm) W buffer layer indicate both heterostructures have PMA, with out-of-plane coercive fields of $H_c \approx 35\,\text{Oe}$ and $H_c \approx 175\,\text{Oe}$, respectively. To unravel the cause of this different magnitudes in coercive field, we perform cross-section high resolution transmission electron microscopy (HR-TEM) imaging on the W(4)/Co-Fe-B(1.4) and W(14)/Co-Fe-B(1.4) heterostructures. As presented in Fig. 1(c) and (d), the HR-TEM results show that the W buffer layer is partially amorphous for the thin W case, whereas it is crystalline (body-centered-cubic) for the thick W case. Therefore, microstructure analysis suggests that the enhanced out-of-plane coercive field for the thick W heterostructure is correlated to the change of W buffer layer texture. This correlation might originate from the change of domain nucleation sites or domain wall pinning sites during the phase transition of W buffer layer. Note that the MgO layers in both samples are not crystalline, which indicates that the templating of Co-Fe-B crystallization during annealing process is mainly affected by the W buffer layer, in contrast to the previously-studied Co-Fe-B/MgO templating effect [37,38]. Nevertheless, all annealed W($t_\text{W}$)/Co-Fe-B(1.4) heterostructures with $1\,\text{nm} \leq t_\text{W} \leq 16\,\text{nm}$ show PMA and allow us to perform DL-SOT characterization using techniques that are suitable for devices with PMA.

**B. Hysteresis loop shift measurements**

To systematically characterize the DL-SOT efficiency from W/Co-Fe-B heterostructures, we perform the current-induced hysteresis loop shift measurements [27,31,36] on patterned Hall-bar devices, as shown in Fig. 2(a). In this type of measurement, we sweep out-of-plane field $H_z$ to obtain hysteresis loops from anomalous Hall voltage under the influence of an extra in-plane bias field $H_x$. The purpose of $H_x$ is to overcome the effective field originating from interfacial Dzyaloshinskii-Moriya interaction (DMI) $H_{DMI}$ and to realign domain wall moments [39,40]. When $|H_x| \geq H_{DMI}$, the current-induced DL-SOT acting on the domain wall moments can be observed as an out-of-plane effective field $H_{eff}^z$, which will lead to a shift in the out-of-plane hysteresis loop. Therefore, by obtaining the hysteresis loop shifts under different applied currents ($I_{DC}$) as shown in Fig. 2(b), the DL-SOT strength can be quantified. We summarize the $I_{DC}$ dependence of switching fields and $H_{eff}^z$ for W(4)/Co-Fe-B(1.4) and W(15)/Co-Fe-B(1.4) samples in Fig. 2(c) and (d), respectively. It is obvious that for both cases $H_{eff}^z$ depends on the applied $I_{DC}$ linearly, which is consistent with previous studies [27,31,36]. However, the slope of $H_{eff}^z$-to-$I_{DC}$ is greater in W(4)/Co-Fe-B(1.4) than in W(15)/Co-Fe-B(1.4), suggesting a more significant DL-SOT effect in the thin W case. The DL-SOT efficiency $\xi_{DL}$ can be quantitatively estimated from the ratio between $H_{eff}^z$ and the current density $J_e$ in the spin-Hall buffer layer by [27,41]

$$\xi_{DL} = \frac{2e}{\hbar}\left(\frac{2}{\pi}\right)\mu_0 M_s t_{FM}^{eff}\left(\frac{H_{eff}^z}{J_e}\right), \qquad (1)$$

where $M_s$ and $t_{FM}^{eff}$ represent saturation magnetization and effective thickness of the FM layer, respectively. $t_{FM}^{eff}$ is the nominal FM thickness subtracted by the magnetically dead layer thickness. These parameters for Co-Fe-B are characterized by vibrating sample magnetometer (VSM) to be $M_s \approx 1500\,\text{emu/cm}^3$ ($1.5\times10^6$ A/m in SI units) and $t_{Co\text{-}Fe\text{-}B}^{eff} \approx 0.7\,\text{nm}$. Since the applied charge current will be flowing in both W and Co-Fe-B layers, $J_e$ in the W layer is calculated by $J_e = J_{DC} \cdot \left[ \rho_{Co\text{-}Fe\text{-}B} t_W / (\rho_{Co\text{-}Fe\text{-}B} t_W + \rho_W t_{Co\text{-}Fe\text{-}B}) \right]$, where $J_{DC} = I_{DC}/(w \cdot t_W)$ is the nominal current density with $w = 5\,\mu\text{m}$ (width of the device). $\rho_{Co\text{-}Fe\text{-}B}$ and $\rho_W$ are the resistivities of Co-Fe-B and W layers, respectively. Therefore, the resulting equation to estimate $\xi_{DL}$ from measured $H_{eff}^z / I_{DC}$ can be expressed as

$$\xi_{DL} = \frac{2e}{\hbar}\left(\frac{2}{\pi}\right)\mu_0 M_s t_{FM}^{eff} w t_W \left(\frac{\rho_{Co\text{-}Fe\text{-}B} t_W + \rho_W t_{Co\text{-}Fe\text{-}B}}{\rho_{Co\text{-}Fe\text{-}B} t_W}\right)\left(\frac{H_{eff}^z}{I_{DC}}\right). \quad (2)$$

The estimated magnitudes of DL-SOT efficiency $|\xi_{DL}|$ as well as relevant magnetic and electrical properties as functions of W buffer layer thickness are summarized in Fig. 3. The W buffer layer thickness dependence of out-of-plane coercive field $H_c$ (Fig. 3(a)) again verifies the microstructure evolution from amorphous to crystalline phase mentioned earlier. Resistance (Fig. 3(b)) and inverse of resistance (Fig. 3(c)) of the devices also indicate the existence of two regimes: resistive thin W ($t_W \leq 4$ nm) and conductive thick W ($t_W \geq 4$ nm). The corresponding resistivities of W buffer layer in these two regimes are estimated to be $\rho_W^{\text{thin (amorphous)}} \approx 185.7\,\mu\Omega\,\text{cm}$ and $\rho_W^{\text{thick (crystalline)}} \approx 90.9\,\mu\Omega\,\text{cm}$. The resulting DL-SOT efficiency $|\xi_{DL}|$ is strongly phase dependent, as shown in Fig. 3(d). For $t_W \leq 4$ nm, $|\xi_{DL}|$ increases while increasing W thickness and can be well-fitted to a spin diffusion model $|\xi_{DL}(t_W)| = |\xi_{DL}^{\text{thin(amorphous)}}|\left[1 - \text{sech}(t_W/\lambda_s^W)\right]$ [4] with $|\xi_{DL}^{\text{thin(amorphous)}}| \approx 0.144$ and spin diffusion length $\lambda_s^{\text{thin(amorphous) W}} \approx 0.9$ nm. In contrast, for all $t_W \geq 4$ nm samples $|\xi_{DL}| \approx 0.03$, which suggests that $|\xi_{DL}^{\text{thick(crystalline)}}| \approx 0.03$ and $\lambda_s^{\text{thick(crystalline) W}} \leq 1.5$ nm. This phase dependence of $|\xi_{DL}|$ in W/Co-Fe-B heterostructures is consistent with previous reports [6,22,26]. However, in this work we further quantitatively estimate the DL-SOT efficiency from crystalline W buffer layers to be $|\xi_{DL}^{\text{thick(crystalline)}}| \approx 0.03$ in W/Co-Fe-B heterostructures. Also note that for all W/Co-Fe-B heterostructures, the magnitude of DMI effective field is estimated to be $|H_{\text{DMI}}| \leq 100\,\text{Oe}$.

**C. Demonstration of spin-orbit torque switching**

In order to demonstrate current-induced DL-SOT switching in these heterostructures, as shown in Fig. 4(a), we apply an in-plane bias field $H_x = 80\,\text{Oe}$ to realign magnetic domain wall moments [39] to facilitate domain expansion. A pulse of charge current with amplitude $I_{sw}$ and pulse width 50 ms is sent into the device to generate a pulsed spin current from the SHE of W buffer layer. The resulting SHE-induced DL-SOT acting on the adjacent Co-Fe-B layer will switch the magnetization when $I_{sw}$ reaches a critical value. As shown in Fig. 4(b), a representative W(4)/Co-Fe-B(1.4) device can be reversibly switched between two magnetization states by critical switching current of $\sim 0.6\,\text{mA}$, which corresponds to critical switching current density $J_c \approx 2.5 \times 10^{10}\,\text{A/m}^2$. Although this value is much lower than the previously reported $J_c \approx 1.6 \times 10^{11}\,\text{A/m}^2$ in a similar heterostructure [18], the result is not unexpected: Since DL-SOT switching in micron-sized samples is mainly governed by domain nucleation and domain wall propagation processes, a lower coercive field sample (in our case $H_c \approx 30\,\text{Oe}$) will give rise to a lower $J_c$ [42]. The feasibility of switching W/Co-Fe-B devices through DL-SOT mechanism can also be predicted from Fig. 3(c) and (d), by viewing them as switching phase diagrams. Based on this concept, we can predict that within the range of $I_{sw}$ we send into devices ($|I_{sw}| \leq 10\,\text{mA}$), no DL-SOT switching will be observed from the W(15)/Co-Fe-B(1.4) sample. Indeed, we only observe successful current-induced DL-SOT switching in thin (amorphous) W samples but not in thick (crystalline) samples.

## IV. RESULTS FROM W/FERRIMAGNETIC BILAYERS

### A. DL-SOT characterization

Recently, it has also been shown that SOT from 5d transition metals (Ta and Pt) or topological insulators ($Bi_2Se_3$) can be utilized to control magnetic moments in a ferrimagnetic layer such as Co-Tb and Co-Gd [31-33,43]. However, a detailed study on DL-SOT from W/ferrimagnet heterostructures has yet to be reported. We employ the same characterization procedure for W($t_W$)/Co-Fe-B(1.4) heterostructures on our W($t_W$)/Co-Tb(10) heterostructures. As shown in Fig. 5(a), the $Co_{63}Tb_{37}(10)$ sputtered onto W(3) buffer layer shows bulk PMA in its as-deposited state, with $H_c \approx 350$ Oe, which is greater than the $H_c$'s of W/Co-Fe-B films. Representative HR-TEM result, as shown in Fig. 5(b), further indicates that both W(3) and Co-Tb(10) layers are amorphous. Representative hysteresis loop shift measurement results from W(3)/Co-Tb(10) and W(10)/Co-Tb(10) Hall-bar devices are shown in Fig. 5(c) and (d), respectively. The $H_{\text{eff}}^z / I_{DC}$ ratios are estimated to be -3.3 Oe/mA for W(3)/Co-Tb(10) and -0.7 Oe/mA for W(10)/Co-Tb(10). The $H_{\text{eff}}^z / I_{DC}$ difference between these two samples again suggests W phase dependence of the SHE. We further estimate the DL-SOT efficiencies of these two samples with different W thickness: $|\xi_{DL}(t_W = 3 \text{ nm})| \approx 0.10$ and $|\xi_{DL}(t_W = 10 \text{ nm})| \approx 0.026$. To calculate $|\xi_{DL}|$ for W/Co-Tb samples from Eqn. (2), $M_s \approx 265$ emu/cm$^3$ and $\rho_{\text{Co-Tb}} = 200 \ \mu\Omega$ cm of the Co-Tb layer are separately characterized by VSM and four-point measurements. No obvious magnetic dead layer is found in the as-deposited W/Co-Tb heterostructures.

We summarize the W thickness dependence of out-of-plane field $H_c$ and DL-SOT efficiency $|\xi_{DL}|$ for W($t_W$)/Co-Tb(10) in Fig. 6(a) and (b), respectively. While no obvious W thickness dependent trend is found for $H_c$ (since the PMA for Co-Tb has a bulk origin), the behavior of $|\xi_{DL}|$ for W/Co-Tb resembles that found in the W/Co-Fe-B system. The thickness dependence of $|\xi_{DL}|$ again can be fit into a spin-diffusion model and we are able to extract $|\xi_{DL}^{\text{thin(amorphous)}}| \approx 0.116$ and $\lambda_s^{\text{thin(amorphous) W}} \approx 1.1$ nm for thin W regime. For thick W regime, $|\xi_{DL}^{\text{thick(crystalline)}}| \approx 0.026$ and $\lambda_s^{\text{thick(crystalline) W}} \leq 2$ nm. Note that the estimated $|\xi_{DL}|$'s for W/Co-Tb are all slightly smaller but fairly close to the estimated $|\xi_{DL}|$'s for W/Co-Fe-B heterostructures, which means that the spin transparency factor $T_{\text{int}}^{\text{W/Co-Tb}} \leq T_{\text{int}}^{\text{W/Co-Fe-B}}$, with the assumption that $\theta_{SH}^{W}$ have the same phase dependent trend in both cases. It is possible that the amorphous nature of W/Co-Tb interface leads to a smaller spin transparency factor while compare to the (partially) crystalline W/Co-Fe-B interface. A similar result of smaller spin transparency factor has also been reported by Finley and Liu for Ta/Co-Tb system (while compare to Ta/Co-Fe-B) [31]. In contrast, the magnitude of DMI effective fields are estimated to be $|H_{\text{DMI}}| \approx 250\,\text{Oe}$ for W/Co-Tb samples, which is greater than those in W/Co-Fe-B heterostructures. More detailed studies on Co-Tb domain wall structure and domain expansion dynamics might elucidate the difference of DMI strength in these two systems.

### B. DL-SOT switching

Note that current-induced DL-SOT switching cannot be realized in W($t_W$)/Co-Tb(10) devices, as can be predicted by the switching phase diagrams Fig. 5(c) and (d). For example, $|I_{sw}| > 10\,\text{mA}$ is required to see possible current-induced switching in the W(3)/Co-Tb(10) device, but a current this large will typically destroy our samples. To achieve DL-SOT switching, we further reduce the thickness of deposited Co-Tb from 10 nm to 3.5 nm. In this thin Co-Tb case, the coercive field of Co-Tb is reduced to $H_c \approx 10\,\text{Oe}$, which is beneficial for observing magnetization switching due to a lower depinning field. In Fig. 6(d), we show a representative current-induced DL-SOT switching curve from a W(4)/Co-Tb(3.5) Hall-bar device with lateral dimensions of $10\,\mu\text{m} \times 60\,\mu\text{m}$. The critical switching current is of ~1 mA, from which the critical switching current density is estimated to be $J_c \approx 1.4 \times 10^{10}$ A/m$^2$. This number is even smaller than the case of W/Co-Fe-B, mainly due to the smaller $H_c$ and magnetization $M_s t_{FM}^{eff}$ of thinner Co-Tb layer. Overall, our results indicate that current-induced DL-SOT in W/Co-Tb heterostructure is an efficient mechanism to induce magnetization dynamics therein. Although previous study had shown that the field-like component of SOT could also possibly exist in a TM/Co-Tb system [32], the magnitude is much smaller than its dampinglike counterpart. Therefore, we conclude that the magnetization switching we observe here is mainly due to current-induced DL-SOT from the SHE of amorphous W.

## V. CONCLUSION

To summarize, we show that current-induced DL-SOT efficiencies $|\xi_{DL}|$ from both W/Co-Fe-B (TM/ferromagnetic) and W/Co-Tb (TM/ferrimagnetic) heterostructures depend on the microstructure of W buffer layer. Through hysteresis loop shift measurements, we estimate $|\xi_{DL}^{thin(amorphous)}| \approx 0.116 - 0.144$ for magnetic heterostructures with thin, amorphous W, while $|\xi_{DL}^{thick(crystalline)}| \approx 0.026 - 0.030$ for heterostructures with thick, crystalline W. By comparing results from both systems, we find the spin transparency factor $T_{int}^{W/Co-Tb} \leq T_{int}^{W/Co-Fe-B}$. We further demonstrate current-induced DL-SOT switching in both W/Co-Fe-B and W/Co-Tb heterostructure systems. Our comparative studies therefore suggest that both ferromagnetic and ferrimagnetic layers can be controlled by the SHE of W, and the current-induced loop shift technique can not only be utilized to quantitatively determine the DL-SOT efficiencies, but also be employed to predict the feasibility of current-induced DL-SOT switching from various magnetic heterostructures.

## ACKNOWLEDGMENTS

This work is supported by the Ministry of Science and Technology of Taiwan (MOST) under Grant No. MOST 105-2112-M-002-007-MY3.

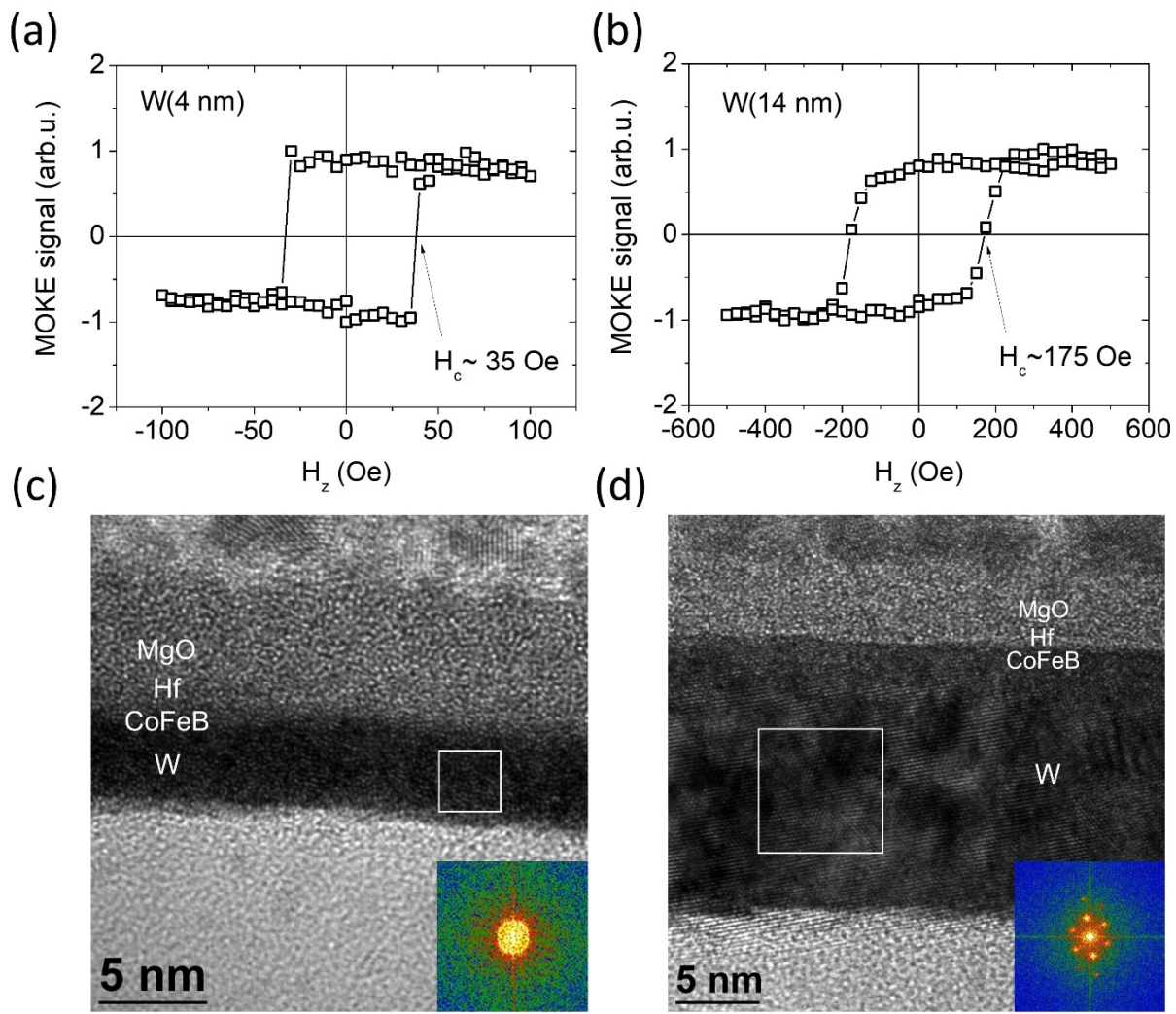

Figure 1. Out-of-plane hysteresis loops of (a) W(4)/Co-Fe-B(1.4)/Hf(0.5)/MgO(2) and (b) W(14)/Co-Fe-B(1.4)/Hf(0.5)/MgO(2) magnetic heterostructures. Cross section HR-TEM imaging results from (c) W(4)/Co-Fe-B(1.4)/Hf(0.5)/MgO(2) and (d) W(14)/Co-Fe-B(1.4)/Hf(0.5)/MgO(2) magnetic heterostructures. The subpanels are the diffractograms derived by reduced fast Fourier transformation (FFT) from the regions of interests (white boxes).

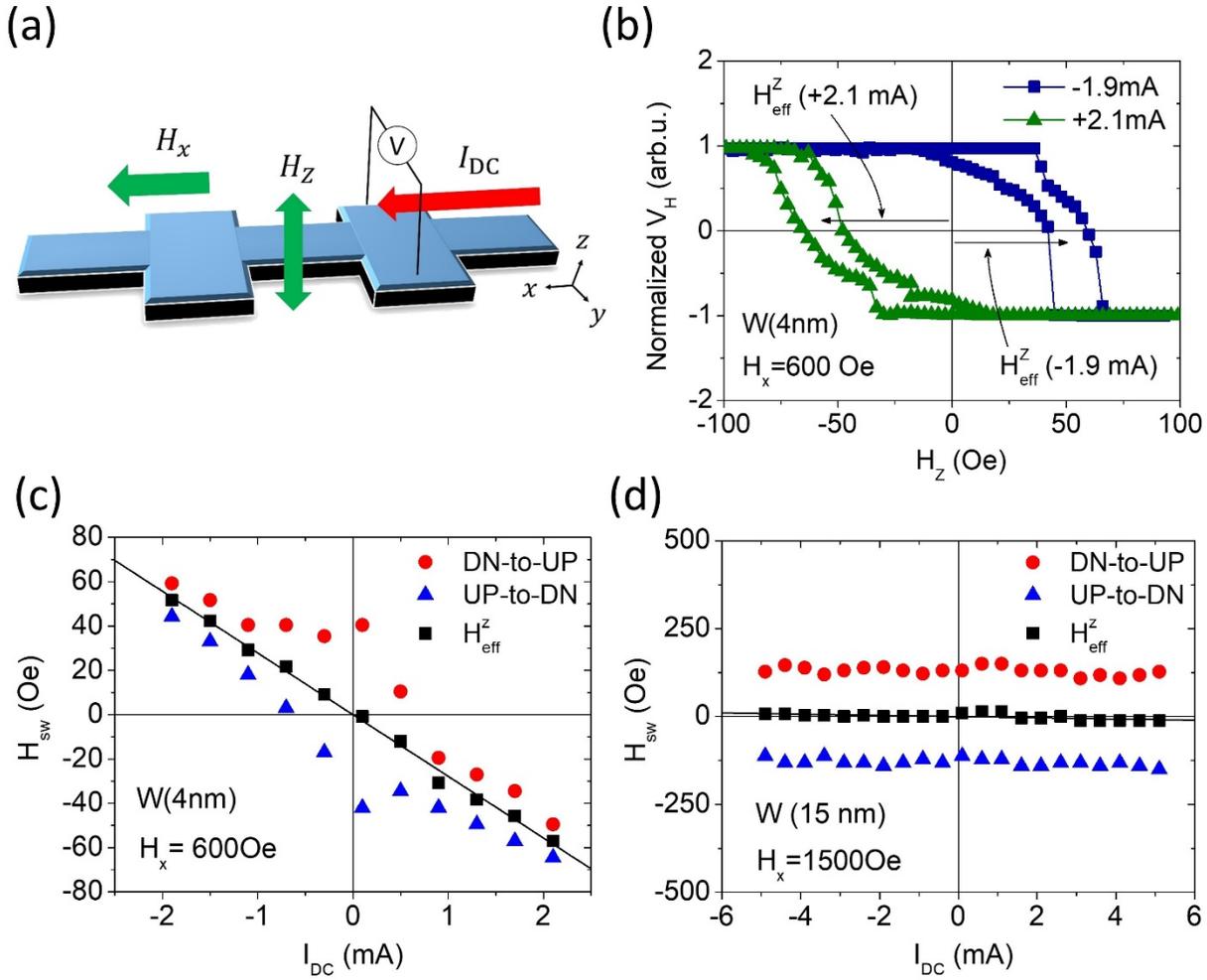

Figure 2. (a) Schematic illustration of anomalous Hall voltage measurement. (b) Representative shifted Hall voltage loops from a W(4)/Co-Fe-B(1.4) sample with different DC currents $I_{DC}$ and an in-plane bias field $H_x = 600$ Oe. (c,d) Switching fields $H_{sw}$ of W(4)/Co-Fe-B(1.4) and W(15)/Co-Fe-B(1.4) samples for down-to-up (red circles) and up-to-down (blue triangles) switching processes as functions of $I_{DC}$, with $H_x = 600$ Oe and 1500 Oe, respectively. $H_{\text{eff}}^z$ (black squares) represent the center of Hall voltage loops. The solid lines represent linear fits to $H_{\text{eff}}^z$ data.

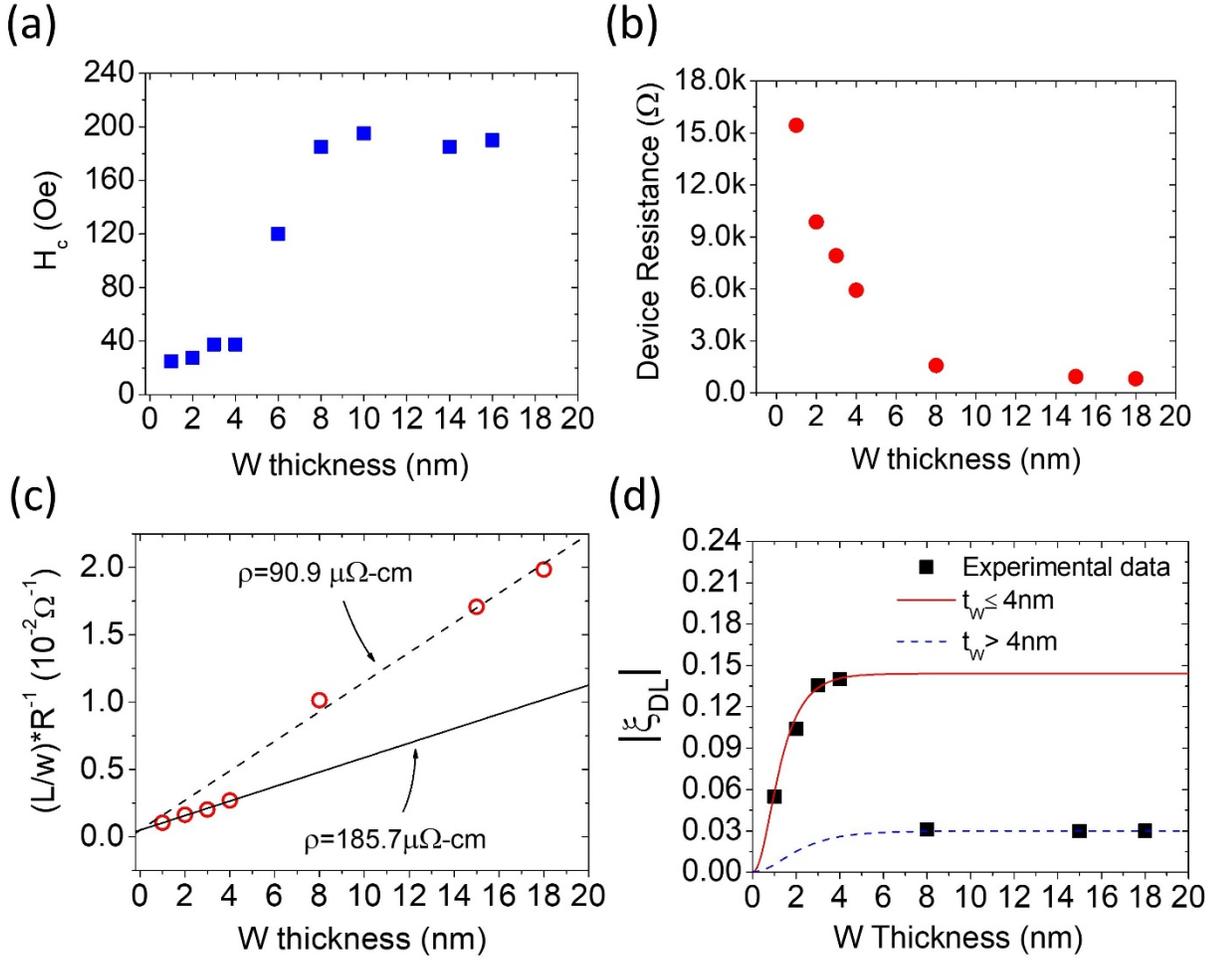

Figure 3. (a) Out-of-plane coercive field $H_c$ of the Co-Fe-B layer, (b) Hall-bar device resistance, (c) inverse of the Hall-bar device resistance, and (d) the magnitude of DL-SOT efficiency $|\xi_{DL}|$ of W/Co-Fe-B magnetic heterostructures as functions of W thickness ($t_W$). $L$ and $w$ in (c) stand for length and width of the Hall-bar device channel, respectively. The solid line and dashed line in (c) represent linear fits to $t_W \leq 4\,\text{nm}$ and $t_W > 4\,\text{nm}$ data, respectively. The red solid line and blue dashed line in (d) represent fits to a spin diffusion model for $t_W \leq 4\,\text{nm}$ and $t_W > 4\,\text{nm}$ data, respectively.

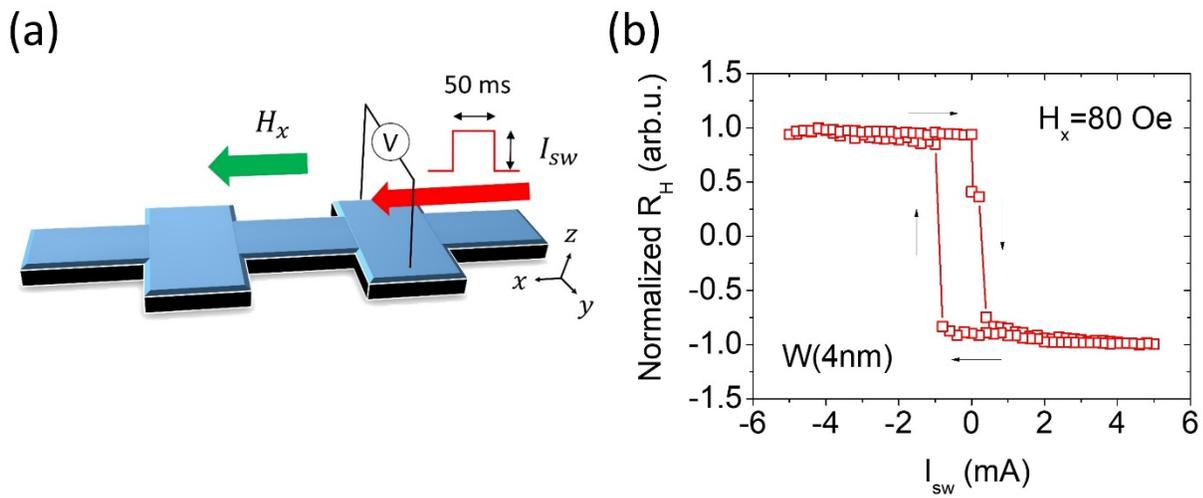

Figure 4. (a) Schematic illustration of current-induced SOT switching measurement. $I_{sw}$ represents the amplitude of injected current pulse. The applied current pulse duration is 50 ms. (b) A representative current-induced SOT switching result from a W(4)/Co-Fe-B(1.4) Hall-bar sample under in-plane bias field $H_x = 80\,\text{Oe}$. The black arrows represent the sweeping directions of applied current pulse $I_{sw}$.

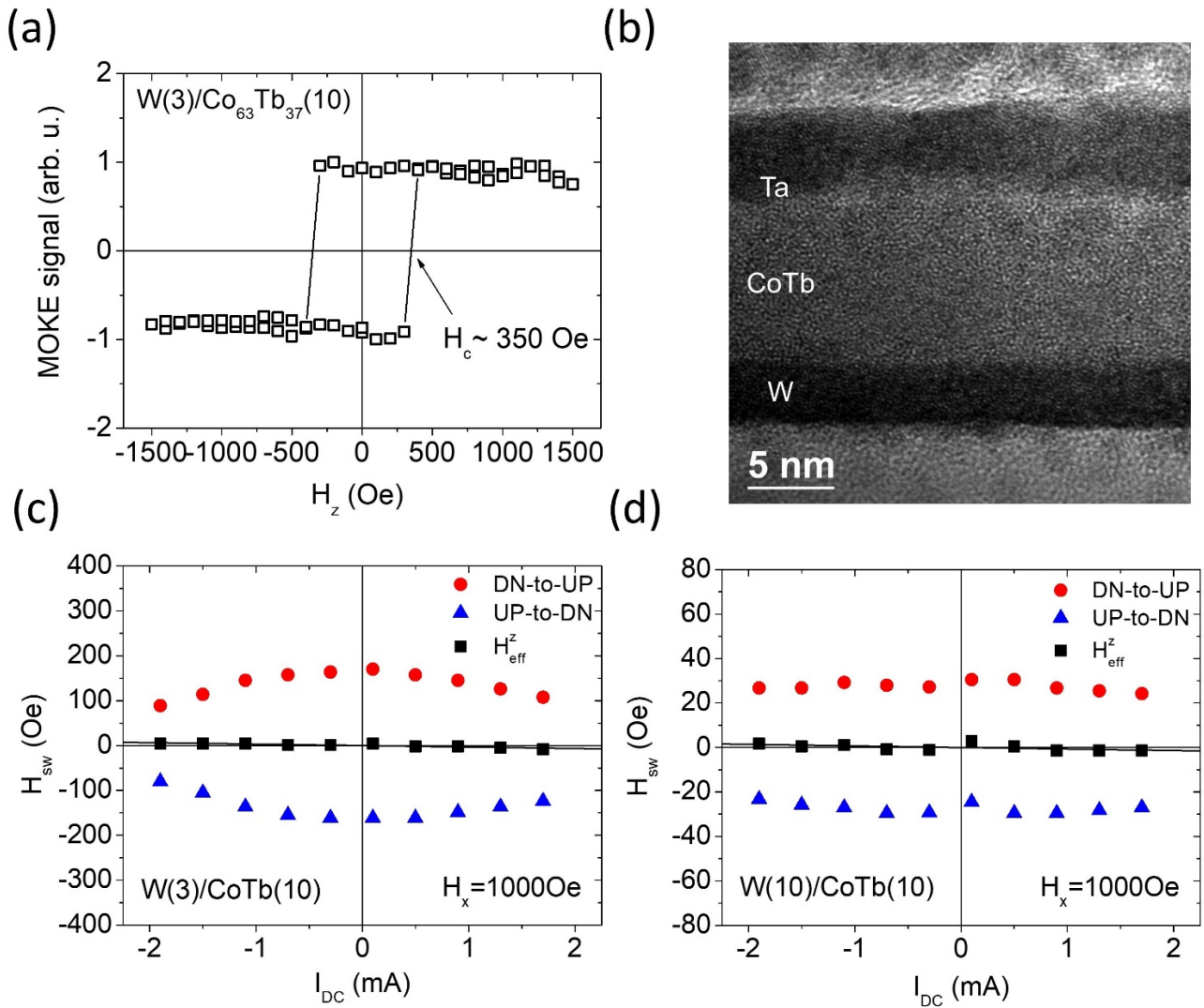

Figure 5. (a) Representative out-of-plane hysteresis loop of a W(3)/Co-Tb(10) heterostructure. (b) Cross section HR-TEM imaging result from a W(3)/Co-Tb(10) sample. (c,d) Switching fields $H_{sw}$ of W(3)/Co-Tb(10) and W(10)/Co-Tb(10) samples for down-to-up (red circles) and up-to-down (blue triangles) switching processes as functions of $I_{DC}$, both with $H_x = 1000$ Oe. $H_{eff}^z$ (black squares) represent the center of Hall voltage loops. The solid lines represent linear fits to $H_{eff}^z$ data.

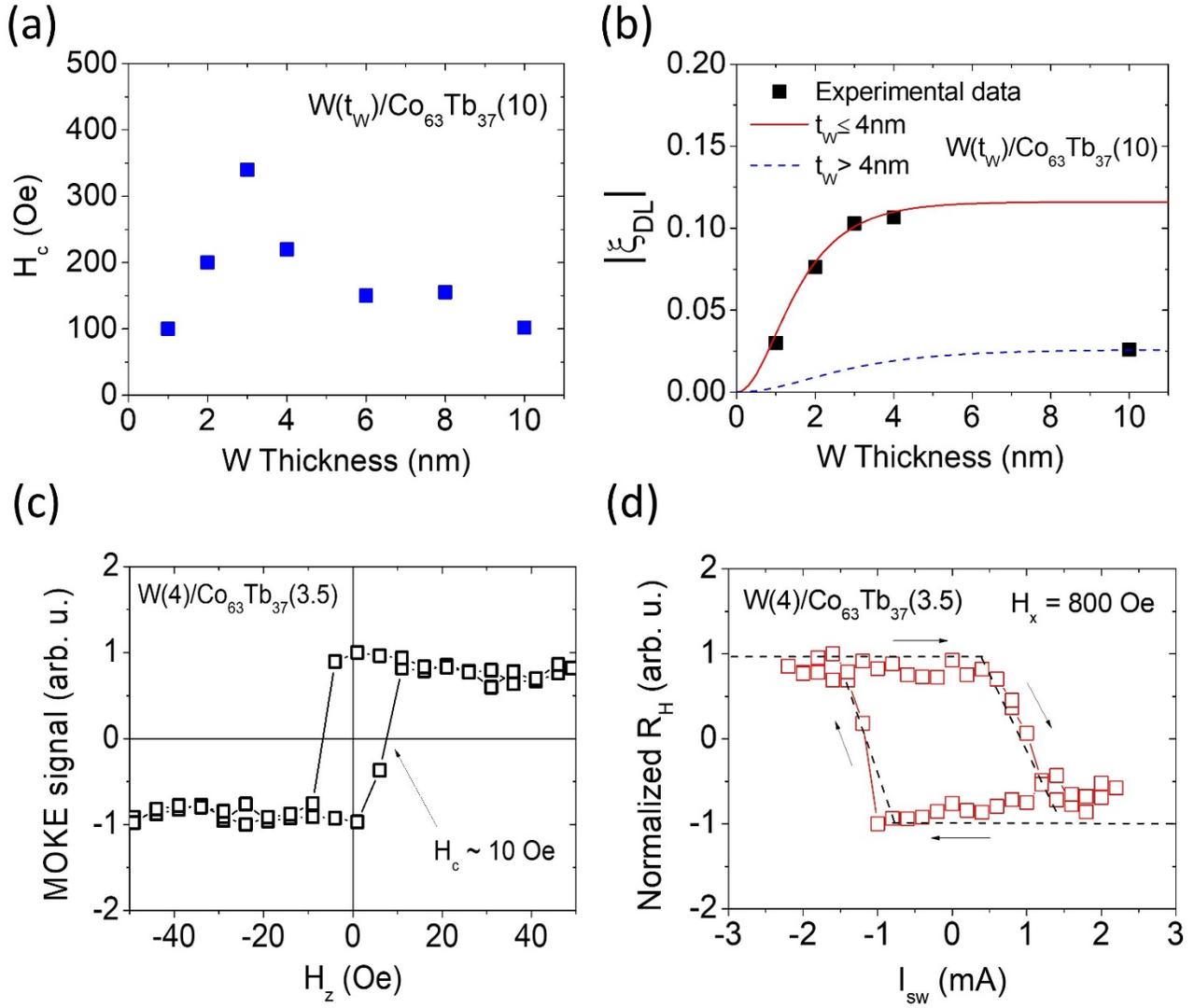

Figure 6. (a) Out-of-plane coercive field $H_c$ of the Co-Tb layer and (b) the magnitude of DL-SOT efficiency $|\xi_{DL}|$ of W($t_W$)/Co-Tb(10) magnetic heterostructures as functions of W thickness ($t_W$). The red solid line and blue dashed line in (b) represent fits to a spin diffusion model for $t_W \leq 4\,\text{nm}$ and $t_W > 4\,\text{nm}$ data, respectively. (c) Out-of-plane hysteresis loop of a W(4)/Co-Tb(3.5) heterostructure. (d) Current-induced SOT switching curve of a W(4)/Co-Tb(3.5) Hall-bar sample under in-plane bias field $H_x = 800\,\text{Oe}$. The black arrows represent the sweeping directions of applied current pulse $I_{sw}$. The dashed lines serve as guide to the eye.